\documentstyle[epsf]{myeuro}

\def\etal{{\hbox{{\tenit\ et al.\/}\tenrm :\ }}}

\def\And{{\rm and\ }}

\newif\ifboo \boofalse

\def\Review#1{\boofalse{\it #1},}
\def\Name#1{{\sc #1},}
\def\Vol#1{\ifboo Vol. {\bf #1}\else{\bf #1}\fi}
\def\Year#1{\ifboo #1\else(#1)\fi}

\def\Page#1{\ifboo {\rm p. #1}\else{\rm #1}\fi}

\def\prl{Phys. Rev. Lett.}

\def\euro{Europhys. Lett.}

\def\pr {Phys. Rev.}

\def\sss                               {
     \scriptscriptstyle                       }

\def\e{\epsilon}

\font\tenrm = cmr10
\font\tenit = cmsl10
\begin{document}
\def\equote#1#2#3#4#5{\Name{#1} \Review{#2} \Vol{#3} \Year{#4} \Page{#5}.} 
\def\qquote#1#2#3#4#5{\Name{#1} \Review{#2} \Vol{#3} \Year{#4} \Page{#5};}  
\def\nquote#1#2#3#4#5{\Name{#1} \Review{#2} \Vol{#3} \Year{#4} \Page{#5}}  
\def\preprint#1#2{\Name{#1} \Review{#2}}  
\def\qbook#1#2#3{{ #1,\ in {\sl#2}, edited by #3}.\par}
\def\nbook#1#2#3{{ #1,\ in {\sl#2}, edited by #3}}
\def\bbook#1#2#3{{ #1,\ in {\sl#2}, edited by #3}.}
\def\trans#1#2#3{[ {\sl #1} {\bf #2},\ #3 ]}
\shorttitle{E. Shopen and Y. Meir, Enhancement of quantum-dot peak-spacing fluctuations etc.}
\title{Enhancement of quantum dot peak-spacing fluctuations in the fractional quantum Hall regime}
\author{Elad Shopen and Yigal Meir}
\institute{Department of Physics,
 Ben-Gurion University, Beer Sheva 84105, ISRAEL}
\pacs{
\Pacs{73}{61.$-$r\ }{multilayers, superlattices, quantum wells,
wires, and dots}
\Pacs{73}{40.Hm\ }{Quantum Hall effect (integer and fractional)}
\Pacs{74}{40.$+$k\ }{Fluctuations}
}

\maketitle
\begin{abstract}
The fluctuations in the spacing of the tunneling resonances through a quantum
dot have been studied in the quantum Hall regime. Using the fact that the
ground-state of the system is described very well by the Laughlin wavefunction, 
we were able to determine accurately,  via classical  Monte Carlo calculations, 
 the amplitude and distribution of the peak-spacing fluctuations.
  Our results clearly
 demonstrate a big enhancement of the fluctuations as the importance of the
 electronic correlations increases,  namely as the density decreases and
filling factor becomes
 smaller. We also find that the distribution of the fluctuations approaches a Gaussian with
increasing density of random potentials.  
\end{abstract}

Several experimental studies\cite{sivan,simmel,patel} have recently demonstrated 
that the fluctuations in the ground-state energy of a quantum dot,  which 
are manifested in the fluctuations in the resonant-tunneling-peak spacings, 
are much larger than what one expect from models that ignore electron
 correlations.
Numerical studies\cite{sivan,prus,koulakov,blanter}
  have indeed revealed an enhancement of the ground-state
energy fluctuations due to electron-electron interactions.

In this work we present calculations for an interacting electron system in
a regime that can be treated almost exactly - the quantum Hall regime. The
ground-state wavefunction in this regime is faithfully described  by the
Laughlin wavefunction \cite{laughlin}. Consequently,  as long as the
potential fluctuations do not mix in excited states (i.e. when the potential
energy is smaller than the gap),  the peak-spacing fluctuations (PSF) 
can be evaluated as
expectation values in the Laughlin state. As such expectation values can be 
easily calculated using the plasma analogy \cite{laughlin},  via, e.g., classical
Monte Carlo simulations,  we are able to obtain the magnitude of the PSF,
  their distribution and their dependence on the range of the potentials
and the electron number. Indeed we find that the more important the electronic
correlations (the lower the filling factor),  the larger the magnitude of
 the PSF. In addition,  we also find that the
  distribution of PSF is Gaussian,  in agreement with
  experiments \cite{sivan,simmel,patel}.

The experimentally measured quantity is the spacings between the 
resonant-tunneling peaks through the quantum dot. At low enough temperatures 
(smaller
than the excitation energies of the dot),  the peak spacing is determined
by the addition spectrum,   
\begin{equation}
\Delta_2^{\sss (N)} \equiv \left( E_g^{\sss (N+1)}- E_g^{\sss (N)}\right) - 
\left( E_g^{\sss (N)}- E_g^{\sss (N-1)}\right) , 
\label{eq:Eg}
\end{equation}
where $E_g^{\sss (N)}$ is the ground-state energy of the N-particle system.
In the constant-interaction model \cite{kastner}
 $E_g^{\sss (N)}=N(N-1)U/2 + \sum_i^N \e_i$, 
where $U$ is the charging energy and $\e_i$ is the single-electron spectrum.
In this model the peak-spacing fluctuations  are determined by the 
fluctuations in the single-electron spectrum,  which are described by
random-matrix theory,  in contrast with the experimental observations. This
 deviation from random-matrix theory was attributed to the importance of
 electronic correlations,  which are not captured in the constant-interaction
 model.

For strong magnetic fields,  in particular in the fractional quantum Hall regime,  
electronic correlations become very significant and dominate the
underlying physics. The ground-state wavefunction of $N$ electrons in 
a quantum dot in a strong
  magnetic field,  with filling factor $\nu=1/m$,  can be described very well
  by the Laughlin wavefunction \cite{laughlin}
\begin{equation}
\Psi^{(\sss 1/m)}\left(z_1,...,z_N\right) = 
\prod_{i<j}\,(z_i-z_j)^m\ e^{-\sum_i|z_i|^2/4}  , 
\label{eq:laughlin}
\end{equation}
where $z_i$ denotes the complex coordinates of the $i$-th particle,
 $z_i=x_i+iy_i$, 
and all lengths are expressed in units of the magnetic length,
$\ell_H\equiv \sqrt{\hbar c/eH}$.

Since the magnetic field quenches the kinetic energy 
and the interaction energy varies
smoothly with the number of particles, the fluctuations in the peak spacings
 stem only from fluctuations in the potential energy. 
If the energy gap due to correlations is large enough (compared to the potential), 
the potential energy in
 the presence of a random potential $V(z)$,  is given by
$E_{pot}^{\sss (N)} = \int d^2r\rho_{\sss N}^{\sss (1/m)}(r)
V(\stackrel{\rightarrow}{r})$,  
where $\rho_{\sss N}^{\sss (1/m)}(r)$ is the single-particle distribution function, 
\begin{equation}
\rho_{\sss N}^{\sss (1/m)}(r) = \int d^2z_1 \ldots  d^2z_N \ 
|\Psi^{(\sss 1/m)}\left(z_1,...,z_N\right)|^2\ \delta(|z_1|^2-r^2)
\label{eq:rho}
\end{equation}
 
The single-particle distribution $\rho_{\sss N}^{\sss (1/m)}(r)$ can be calculated using
the mapping onto a classical plasma model \cite{laughlin}. This mapping
is based upon rewriting $|\Psi|^2$ as

\begin{equation}
|\Psi^{(\sss 1/m)}\left(z_1,...,z_N\right) |^2 = \exp \left[-\sum_i|z_i|^2/2 +  
m \sum_{i<j} \log(|z_i-z_j|^2) \right] .
\label{eq:plasma}
\end{equation}

Consequently,  expectation values in the ground-state can be expressed as
statistical averages for a classical system of $N$ particles in a harmonic
confining potential and logarithmic interactions. These averages were
evaluated using a classical Monte Carlo approach \cite{metropolis}. 

\begin{figure}
%
\epsfxsize=5in
\epsfbox{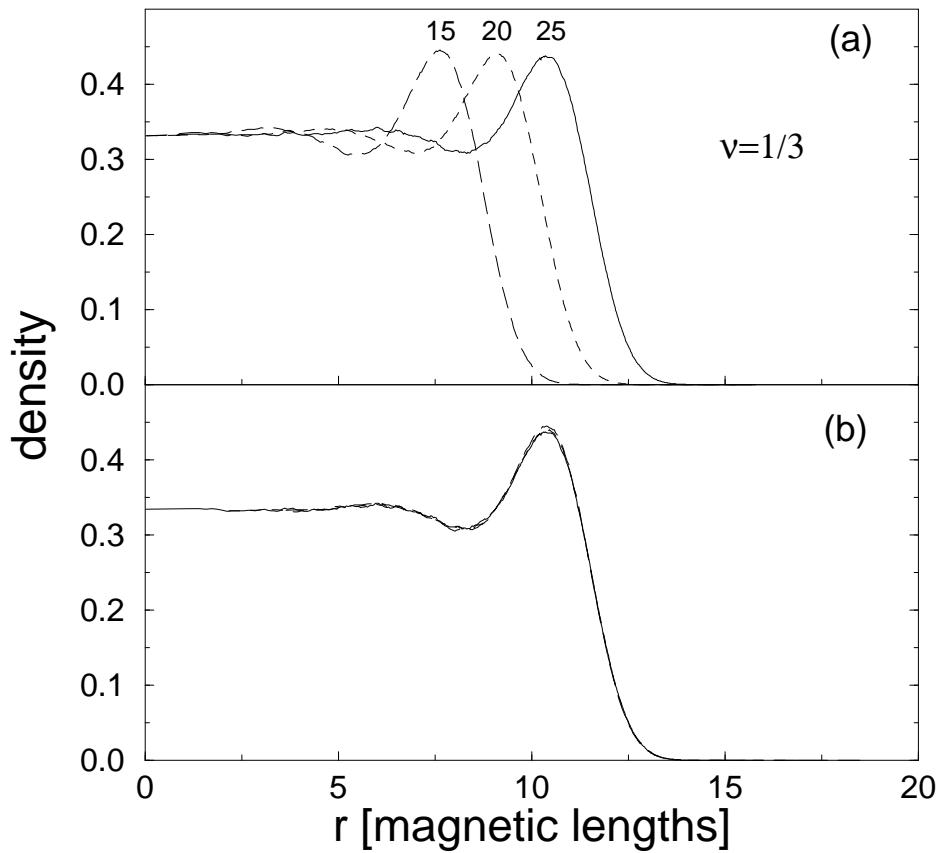}
\caption{(a) Single-particle densities for 15, 20  and 25 particles in the
 $\nu=1/3$ quantum Hall regime. (b) The same densities as in (a), but relatively 
 shifted (see text).}
\label{fig1}
\end{figure}

In fig.~1a we plot the single-particle density of 15, 20 and 25 particles in
the $\nu=1/3$ fractional quantum Hall regime. As can be seen from the figure, 
 the bulk density is constant,  while the edge structure for different
 particle numbers is identical, but  relatively shifted. Since the edge of the
 $N$-particle system in the $\nu=1/m$ case is determined by the maximal
 occupied angular momentum state,  $m(N-1)$, 
 localized at distance $\sqrt{2m(N-1)}$ from the origin, the edge structure of
 $N_1$ and $N_2$-particle systems can be made to collapse by a relative
 shift of $\sqrt{2m(N_1-1)}-\sqrt{2m(N_2-1)}$. In fig.~1b we replot
 the same densities as in fig.~1a,  but relatively shifted in that manner.
   Indeed we see
  that all the distributions collapse onto a single curve.
  This invariance of the edge-structure allows us to deduce immediately
   the $N$-dependence of the peak-spacing fluctuations: writing

\begin{equation}
\Delta_2^{\sss (N)} =  \int d^2r\ 
\Delta_2 \rho_{\sss N}^{\sss (1/m)}(r)\ V(\stackrel{\rightarrow}{r})
\simeq \int d^2r\ 
{{\partial^2 \rho_{\sss N}^{\sss (1/m)}(r)}\over{\partial N^2}}\ 
V(\stackrel{\rightarrow}{r}) , 
\label{eq:delta2}
\end{equation}
(with $\Delta_2 \rho_{\sss N}\equiv \rho_{\sss N+1}+\rho_{\sss N-1}-2
\rho_{\sss N}$), 
and using the abovementioned fact that
\begin{equation}
\rho_{\sss N}^{\sss (1/m)}(r) = 
\rho_{\sss N_0}^{\sss (1/m)}(r+\sqrt{2m(N_0-1)}-\sqrt{2m(N-1)}) , 
\label{eq:rhoN}
\end{equation}
one finds,  after a little algebra,  that for short-range potentials and
 large $N$ 
the fluctuations in $\Delta_2$ scale like 
$<(\Delta_2^{\sss (N)})^2>\sim N^{\sss -3/2}$. Here $<...>$ denotes average
over realizations of the random potential. In particular,
  for delta-function potentials of density $n_0$ and typical strength $V_0$, 
  one finds $<(\Delta_2^{\sss (N)})^2> = A_m n_0 V_0^2 N^{-3/2}$,  where
$A_m$ is $N$-independent for large $N$.
\begin{figure}
%
\epsfxsize=5in
\epsfbox{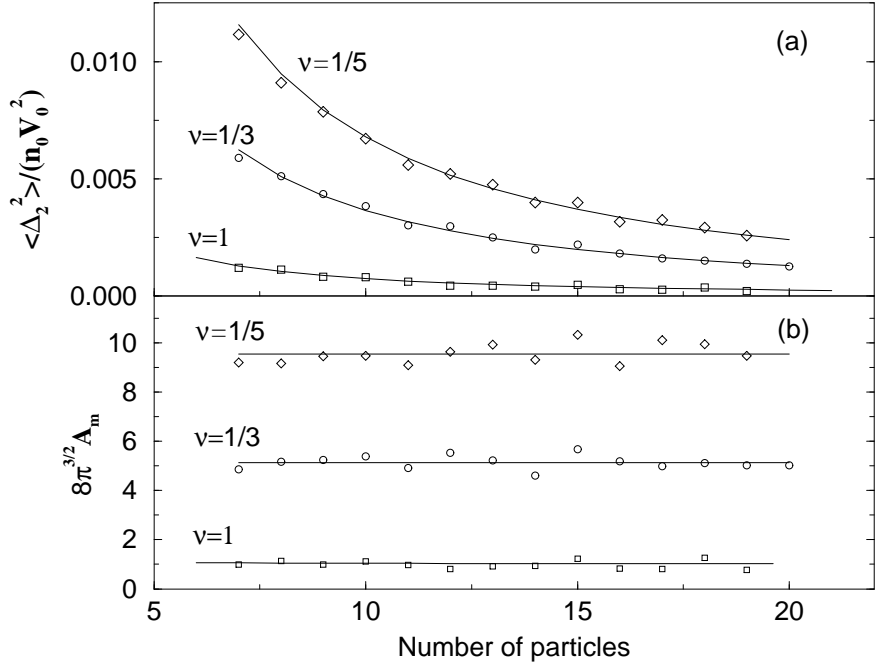}
\caption{(a) Averaged peak-spacing fluctuations for $\nu=1,  1/3$,  and $1/5$.
The continuous lines are fits to $A_m/N^{3/2}$ ($\nu=1/3$ and $1/5$) and 
the exact result for $\nu=1$.   (b) The same data as in (a),  normalized by the
large-$N$ $\nu=1$ exact result.}
\label{fig2}
\end{figure}
In fig.~2a we show the averaged PSF, $<(\Delta_2^{\sss (N)})^2>$, 
 in the $\nu=1,\ 1/3$ and $1/5$ quantum Hall regimes. In agreement
  with the above argument,  the average PSF scale with $N^{-3/2}$
   (continuous curve). 
  In fig.~2b we plot the numerically calculated $A_m$ (scaled by the 
large-$N$ value $A_1\rightarrow1/8\pi^{\sss 3/2}$) as a function of $N$. 
  As expected,  it is indeed $N$-independent within the accuracy
  of the calculation.   The numerically deduced 
  values  lead 
  to 
\begin{eqnarray}
<(\Delta_2^{\sss (N)})^2>_{1/3} / <(\Delta_2^{\sss (N)})^2>_{1}
 &=& 5.0\pm0.3 \nonumber\\
<(\Delta_2^{\sss (N)})^2>_{1/5} / <(\Delta_2^{\sss (N)})^2>_{1}
 &=& 9.2\pm0.5  .
\label{eq:ratios}
\end{eqnarray}
These calculations clearly demonstrate the enhancement of the PSF due to the
increased role of correlations, as the filling factor is lowered
 \cite{density}. (Note that the PSF are independent of the charging
energy in this regime).

As the range of the random potentials, $\sigma$, increases, the PSF
are reduced and the ratios (eq.(\ref{eq:ratios})) decrease towards unity
(fig.~3). However, once $\sigma$ becomes of the order of the size of
the dot, $R\propto\sqrt{N}$,
 the PSF become independent of system size and the ratio
increase again, leading to a nonmonotonic dependence on $\sigma$.
In the  limit
of very smooth potentials, $\sigma\rightarrow\infty$, one can show that the PSF
are given by
\begin{equation}
<(\Delta_2^{\sss (N)})^2> = {5\pi<V_0^2>n_0\over{16\sigma^6}} \left(\int
d^2r\ r^2 \  \Delta_2 \rho_{\sss N}^{\sss (1/m)}(r)\right)^2 \ .
\label{longrange}
\end{equation}
The quantity in parentheses is twice the second
 derivative of the total angular momentum with respect to $N$ and  is equal to $2m$.
 Thus in this limit,  one finds
\begin{eqnarray}
<(\Delta_2^{\sss (N)})^2>_{1/3} / <(\Delta_2^{\sss (N)})^2>_{1}
 &=& 9 \nonumber\\
<(\Delta_2^{\sss (N)})^2>_{1/5} / <(\Delta_2^{\sss (N)})^2>_{1}
 &=& 25  , 
\label{eq:longrangeratios}
\end{eqnarray}
 namely there is an enhancement of the PSF also in this limit.
As the range of the potentials is determined experimentally by the
distance to the donor layer, which is of the same order of the dot
size, the ratio $\sigma/R$ can be varied experimentally in the
relevant regime and  these predictions, including the nonmonotonicity,
could be tested.

\begin{figure}
%
\epsfxsize=5in
\epsfbox{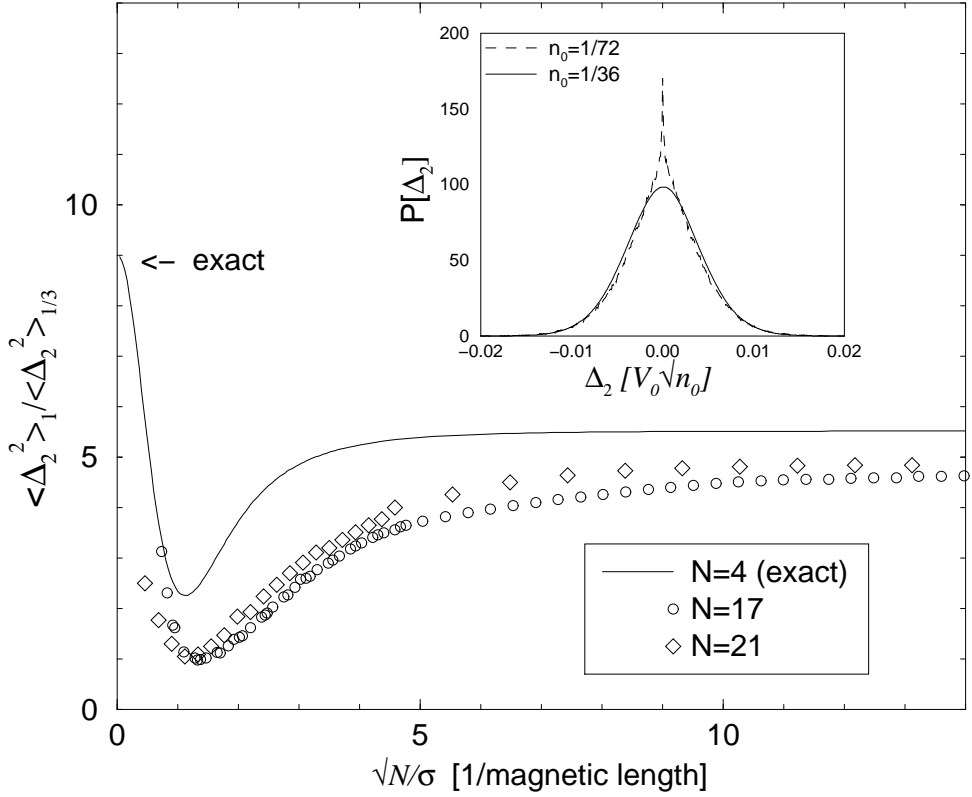}
\caption{
The ratio of the average peak-spacing fluctuations between the $\nu=1$ and the 
 $\nu=1/3$ quantum Hall regimes as a function of the range of the potentials,
 scaled by the dot size $R\propto \sqrt{N}$.
{\sl inset:} The distribution of the peak-spacing fluctuations for two
 densities of random potentials. For the larger densities the
 distribution approaches a Gaussian.}
\label{fig3}
\end{figure}

The observed features can be understood as follows. Based on
Wen's description of the edges of the quantum Hall liquid \cite{wen},
Kinaret et al. have shown that for filling factors $\nu=1/m$, 
$\Delta_2 = mv/R$, where $R$ is the radius of the system and $v$ the edge
velocity. This result can be simply understood  -- when the number
of particles increases by one, the energy of the highest-occupied angular 
momentum state increases by 
$\nabla V \partial R/\partial N$, where $\nabla V$ is the potential
gradient near the edge of the system. Since $R\sim \sqrt{2mN}$ 
this immediately leads to 
$\Delta_2  \sim m \nabla V /R \propto m v/R$. This result implies
$<(\Delta_2)^2> \sim m^2 <(\nabla V)^2> / N$. As $<(\nabla V)^2> \sim 
n <V_0^2> / \sigma^6$, where $n$ is the number of random potentials, one
immediately reproduces eq.(\ref{longrange}).
 As the derivation here
relied on the sharpness of the edge it only strictly applies in that limit.
For finite-range potentials one expects the internal structure of the edge to
modify this result, which in indeed observed numerically.

Lastly we consider the full distribution of the PSF. For simplicity, we 
concentrate on delta-function potentials. In this case the distribution is
 defined by 
\begin{equation}
P(\Delta_2) \equiv \int\ \prod_i d^2r_i\ dV_i\ P(r_i,V_i)\ \ 
 \delta\left(\Delta_2-\sum_i V_i \Delta_2\rho(r_i)\right) , 
\label{distribution}
\end{equation}
where $P(r_i,V_i)$ is the probability of finding a delta-potential of magnitude
$V_i$ at point $r_i$. Assuming a uniform distribution of potentials of density
$n_0$,  and a Gaussian distribution of their amplitudes,  with width $V_0$, 
one finds
\begin{equation}
P(\Delta_2) = \int\ \prod_i {d^2r_i\over A}\ {1\over\sqrt{2\pi\Sigma}}\  
e^{-\Delta_2^2/(2\Sigma^2)} ; \ \ \ 
\Sigma \equiv   V_0^2 \ \sum_i \left(\Delta_2\rho(r_i)\right)^2 .
\end{equation}
When the potentials are dense,  the sum (in the expression for $\Sigma$)
can be approximated by an integral,  and we immediately find that the
$\Delta_2$ is distributed normally with a width that has been calculated
above. Note that this derivation did not involve any information
about the quantum Hall system (except the assumption that one can treat
the potential fluctuations perturbatively),  and is just a manifestation of
the law of big numbers.

When the density of the potentials is lowered,  there is a finite probability
that none of the potentials affects the region of interest -- the edge of the
system. This enhances the probability that $\Delta_2=0$. In the inset of 
fig.~3 we plot
the distribution of the PSF for two different densities of potentials. We 
indeed see that for low enough density there is a sharp peak at $\Delta_2=0$, 
while the distribution approaches a Gaussian for large enough density. We
expect a similar effect when the range of the potential increases.
 
To conclude,  we have demonstrated that the increased importance of the
correlations in the quantum Hall regime leads to an enhancement of the
peak-spacing fluctuations. This can be attributed to the increased rigidity
of the ground-state wavefunction as the filling factor becomes lower.

\section{acknowledgements}
This research was supported by THE ISRAEL SCIENCE FOUNDATION founded 
 by the Israel Academy of Sciences and Humanities - Centers of Excellence
 Program.


\begin{thebibliography}{99}

%
\bibitem{sivan}
\equote{Sivan U.,  \etal}{\prl}{77}{1996}{1123}
%
\bibitem{simmel}
\equote{Simmel F.,  Heinzel T. \And Wharam D. A.}{Europhys. Lett.}
{38}{1997}{123}
%
\bibitem{patel}
\preprint{Patel S. R.,  \etal} {condmat/9708090}.

\bibitem{prus}
\qquote{Prus O.,  \etal}{\pr}{B 54}{1996}{R14289}
\qquote{Berkovits R. \And Sivan U.}{\euro}{41}{1998}{653}
\preprint{Berkovits R.}{condmat/9804107}.

\bibitem{koulakov}
\equote{Koulakov A. A., Pikus F. G. and Shklovskii B. I.}
{\pr}{B 55}{1997}{9223}

\bibitem{blanter}
See,  however,  \qquote{Blanter Ya. M., Mirlin A. D. \And Muzykantskii B. A.}
{\prl}{78}{1997}
{2449}
\preprint{Vallejos R. O., Lewenkopf C. H. \And Mucciolo E. R.}
{condmat/9802124}.

\bibitem{laughlin} 
\equote{R.B. Laughlin}{\prl}{50}{1983}{1395}

\bibitem{kastner}
For a review  see: \equote{M. A. Kastner}{Rev. Mod.
Phys.}{64}{1992}{849}

\bibitem{metropolis}
\equote{M. Metropolis \etal}{J. Chem. Phys.}{21}{1953}{1087}

\bibitem{density}
Throughout this paper we assume that the change in filling factor is caused by 
a change in density (gate-voltage), with constant magnetic field.
 If the magnetic
field is also changed, trivial factors that involve the respective magnetic
lengths will enter the ratios in eqs. (\ref{eq:ratios}) and
 (\ref{eq:longrangeratios}). In fact, if the change in filling factor is fully
caused by a change in the magnetic field, there will be no dependence of the
PSF on the filling factor in the limit of very smooth potentials.

\bibitem{wen}
For a review see \equote{X.-G. Wen}{Int. J. Mod. Phys.}
 {6}{1992}{1711}

\bibitem{kinaret}
\equote{Kinaret J. M.,  \etal}{\pr}{B 46}{1992}{4681}

\end{thebibliography}
\end{document}